\documentclass{PoS}
\usepackage{pifont}
\newcommand{\beqan}{\begin{eqnarray*}}
\newcommand{\eeqan}{\end{eqnarray*}}   
\newcommand{\ba}{\begin{array}}
\newcommand{\ea}{\end{array}}
\newcommand{\no}{\nonumber}

\newcommand\lsim{\mathrel{\rlap{\lower4pt\hbox{\hskip1pt$\sim$}}
    \raise1pt\hbox{$<$}}}
\newcommand\gsim{\mathrel{\rlap{\lower4pt\hbox{\hskip1pt$\sim$}}
    \raise1pt\hbox{$>$}}}

\newcommand{\mrm}{\mathrm}

\title{Status of chiral perturbation theory for light mesons}

\ShortTitle{Mesonic CHPT}

\author{Gerhard Ecker  \\
        University of Vienna, Faculty of Physics, Boltzmanngasse 5,
        A-1090 Wien, Austria\\
        E-mail: \email{gerhard.ecker@univie.ac.at}}

\abstract{The status of chiral perturbation theory in the meson sector
  is illustrated with several topical examples. The longtime
  discrepancy between theory and experiment for the charged pion
  polarizabilities has now been resolved in favour of the chiral
  $SU(2)$ result to next-to-next-to-leading order. For chiral
  $SU(3)$, the main obstacles are the large number of badly known
  coupling constants (LECs) and the lack of convergence of the
  low-energy expansion in many cases of interest. I describe a new
  global fit of the LECs in the strong sector that leads to a
  prediction of the CKM matrix element $V_{us}$ in agreement with the
  latest lattice determinations. The slow convergence of the chiral
  series is particularly acute in transitions with strong final-state
  interactions calling for dispersive treatments. The status of
  dispersive approaches is reviewed for $K_{\ell 4}$ decays and
  for $\eta \to 3 \pi$ decays where precise
  measurements of Dalitz plot distributions have recently become
  available.   }

\FullConference{The 8th International Workshop on Chiral Dynamics, CD2015 \\
		29 June 2015 - 03 July 2015\\
		Pisa,Italy}

\renewcommand{\logo}
{\raisebox{-5mm}
{\parbox{\textwidth}{\begin{flushright}
UWThPh-2015-24\\
October 2015\\
\end{flushright}
\includegraphics{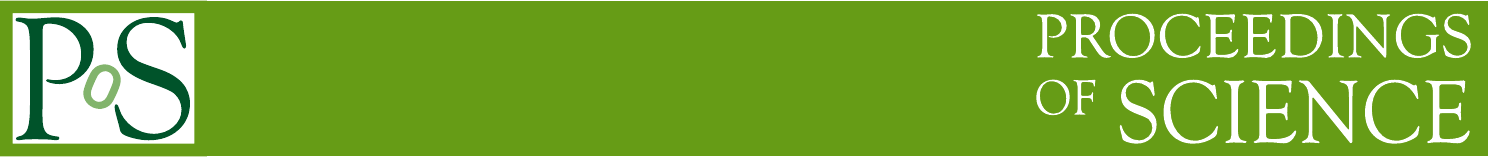}}
}}
\FullConference{{\rm To be published in the Proceedings of \\}
The 8th  International Workshop on Chiral Dynamics \\
  29 June 2015 - 03 July 2015\\
  Pisa, Italy}

\begin{document}

\section{Introduction}
\label{sec:intro}

Chiral perturbation theory (CHPT) in its original form
\cite{Weinberg:1978kz,Gasser:1983yg,Gasser:1984gg} describes the
strong, electromagnetic (external photons) and semileptonic weak
interactions at low energies for the 
light pseudoscalar mesons, pions only for chiral $SU(2)$, the light
pseudoscalar octet for chiral $SU(3)$. Later on, the CHPT scheme was
extended to account also for the nonleptonic weak interactions and for
radiative corrections, requiring the incorporation of dynamical
photons and leptons. A brief review of the relevant chiral Lagrangians
can be found in Ref.~\cite{Bijnens:2014lea}. Schematically, they are
displayed in Table \ref{tab:lecs}.

\begin{table}[!ht]
\label{tab:lecs}
$$
\begin{tabular}{|l|c|} 
\hline
&  \\[-.2cm] 
\hspace{1cm} ${\cal L}_{\rm chiral\; order}$ 
~($\#$ of LECs)  &  loop  ~order \\[8pt] 
\hline 
&  \\
 ${\cal L}_{p^2}(2) $~+~ 
${\cal L}_{p^4}^{\rm odd}(0) $
~+~ ${\cal L}_{G_Fp^2}^{\Delta S=1}(2)$  
~+~ ${\cal L}_{G_8e^2p^0}^{\rm emweak}(1) $ & $L=0$
\\[3pt] 
~+~ ${\cal L}_{e^2p^0}^{\rm em}(1)$ ~+~
${\cal L}_{\mrm{kin}}^{\rm leptons}(0)$ & \\[10pt]
~+~ $\textcolor{red}{{\cal L}_{p^4}(10)}$~+~
${\cal L}_{p^6}^{\rm odd}(23)$
~+~${\cal L}_{G_8p^4}^{\Delta S=1}(22)$
~+~${\cal L}_{G_{27}p^4}^{\Delta S=1}(28)$ &   
$L \le 1$ \\[3pt]
~+~${\cal L}_{G_8e^2p^2}^{\rm emweak}(14)$ 
~+~ ${\cal L}_{e^2p^2}^{\rm em}(13)$
~+~ ${\cal L}_{e^2p^2}^{\rm leptons}(5)$  & \\[10pt] 
~+~ $\textcolor{red}{{\cal L}_{p^6}(90)}$  & $L \le 2$ \\[8pt] 
\hline
\end{tabular}
$$
\caption{Effective chiral Lagrangian in the meson sector for chiral
  $SU(3)$, with the number of LECs in brackets.}
\end{table}

Most of the strong, electromagnetic or semileptonic weak processes
have been calculated up to next-to-next-to-leading order (NNLO), which
includes one- and two-loop contributions. For a review of CHPT at the 
two-loop level see Ref.~\cite{Bijnens:2006zp}, for a repository of
corresponding amplitudes, which is regularly being updated, see
Ref.~\cite{Bijnens:2014gsa}. For the nonleptonic weak interactions and for
radiative corrections, complete calculations exist in general only at
NLO (one-loop level) although attempts at going beyond have been made
in several cases (unitarity corrections, dispersion theory, etc.).

Except for the technical complications of higher orders, the main
obstacle is the rapidly growing number of coupling constants usually
denoted as low-energy constants (LECs),
characteristic for an effective (nonrenormalizable) quantum field
theory. As shown in Table \ref{tab:lecs}, only the strong chiral
Lagrangian has been carried to the NNLO level precisely for that
reason. For instance, for the nonleptonic weak Lagrangian even many
NLO LECs are poorly known. In the strong sector, some progress has
recently been made to estimate many LECs up to NNLO. This will be
reviewed in Sec.~\ref{sec:lecs}.
 
In addition to badly known LECs, the lack of ``convergence'' of the
chiral expansion in many cases is the second major obstacle on the way
to reliable predictions. This refers mostly to calculations in chiral
$SU(3)$ where the natural expansion parameter is $M_K^2/(4\pi F_\pi)^2
\simeq 0.20$. In fact, there are several examples where successive
orders in chiral $SU(3)$ increase by substantially more than 20 $\%$. 
Prominent examples are $\eta \to 3 \pi$ decays that will be
discussed at length in Sec.~\ref{sec:eta}. Instead of calculating
still higher orders in CHPT, the emphasis in the field has shifted
towards dispersive approaches, especially in cases with strong
rescattering in the final state. This status report bears ample
evidence for the growing importance of dispersive treatments in
combination with CHPT.

Although the activity in the field has decreased during the
past 10 years both in theory and experiment, there are several key
experiments still running with a large impact on the field. Two of the
most important recent experimental advances in our field were
presented at this Workshop and will be covered in this talk: the
COMPASS experiment measuring the charged pion polarizabilities and
several recent experiments investigating $\eta \to 3 \pi$ decays.   

Here is an outline of the talk. In Sec.~\ref{sec:lecs} I will
discuss the status of LECs in the strong sector. I will describe our
recent attempts to get a better handle on the LECs of both NLO and
NNLO by means of a global fit yielding a preferred set of LECs
termed BE14. The status of CKM unitarity is reviewed in
Sec.~\ref{sec:ckm}. The longstanding problem of the discrepancy between
theory and experiment for the charged pion polarizabilities has been
resolved recently by the results of the COMPASS experiment. The
history of this problem is briefly recalled in
Sec.~\ref{sec:polar}. Moving up in meson masses, recent activities
in the semileptonic $K_{\ell 4}$ decays are reviewed in
Sec.~\ref{sec:kaon}. To properly describe $\eta \to 3 \pi$ decays
is still a major problem for CHPT, especially in view of recent precise
data for the Dalitz plot distributions. Various dispersive approaches
trying to improve on the NNLO CHPT amplitudes are discussed in
Sec.~\ref{sec:eta}.  

\section{Low-energy constants}
\label{sec:lecs}
\stepcounter{table}

\parbox[c]{0.5\textwidth}{
\begin{flushleft} 
\begin{tabular}{|c|r|r|}
\hline
& & \\[-.3cm] 
  & \hspace*{.3cm} $C_i^r=0$ & Ref.~\cite{Jiang:2015dba} 
\\ \hline 
& & \\[-.3cm] 
$10^3 L_1^r$&   0.67(06)&   0.45(07) \\
$10^3 L_2^r$&   0.17(04)&   0.22(04) \\
$10^3 L_3^r$&$-$1.76(21)& $-$1.66(22)  \\
$10^3 L_4^r$&   0.73(10)& 0.51(12)   \\
$10^3 L_5^r$&   0.65(05)&   2.61(12)  \\
$10^3 L_6^r$&   0.25(09)&  0.73(06)  \\
$10^3L_7^r$&$-$0.17(06)&$-$0.54(05)  \\
$10^3 L_8^r$&   0.22(08)&  1.43(10)  \\
\hline \\[-.3cm]                                                        
$\chi^2$    &   26     & 25   \\
dof         &   9       &  9   \\
\hline
\end{tabular}
\end{flushleft}
}

\vspace*{-6cm}
\hspace*{7cm}  
\parbox[c]{0.5\textwidth}{{\bf Table \arabic{table}:} Fits for NLO
  LECs $L_i^r(\mu)$ for two special cases with fixed NNLO LECs
  $C_i^r(\mu)$. In the second column, all $C_i^r$
  are set to zero. In the third column, we 
  use the $C_i^r$ obtained from a chiral quark model
  \cite{Jiang:2015dba}, itself an update of a previous attempt
  \cite{Jiang:2009uf} to determine the $C_i^r$. The renormalization
  scale is always $\mu=0.77$ GeV.}

\vspace*{3cm}


Chiral perturbation theory as the prototype of an effective field
theory depends on a large number of coupling constants, increasing
rapidly at higher orders. The relevant chiral
Lagrangians in the meson sector, with the associated number of LECs in
brackets, are listed in Table \ref{tab:lecs} for chiral $SU(3)$.

A comprehensive survey of mesonic LECs can be found in
Ref.~\cite{Bijnens:2014lea}. In this talk, I will concentrate on a new
determination of NLO and NNLO LECs in the strong sector
\cite{Bijnens:2014lea} marked in red in Table \ref{tab:lecs}. This new
determination is both an update and an extension of the previous
global fit by Bijnens and Jemos \cite{Bijnens:2011tb}. Referring to
Ref.~\cite{Bijnens:2011tb} for the general strategy of the fit, I
discuss here only the new ingredients of our approach
\cite{Bijnens:2014lea}. 


\begin{itemize} 
\item In addition to the 13 observables 
  employed by Bijnens and Jemos, we have also used the relations
  between the chiral $SU(2)$ LECs $l_j(j=1,\dots,4)$
  \cite{Gasser:1983yg} and the $SU(3)$ LECs $L_i$ \cite{Gasser:1984gg}
  and $C_i$ \cite{Bijnens:1999sh} obtained in
  Ref.~\cite{Gasser:2007sg}. 
\item The altogether 17 input data depend on most of the
  $L_i(i=1,\dots,8)$ and on 34 combinations of the $C_i$. It is
  therefore obvious that one has to make some assumptions about the
  $C_i$ in order to proceed. That this is a nontrivial task
  becomes evident from Table \arabic{table} where we consider two
  special cases with fixed values for the  $C_i$.     
\end{itemize} 
 
Both choices for the $C_i^r$ in Table \arabic{table} clearly lead to
unacceptable fits. In addition to the large values of  $\chi^2$, 
the LECs $L_4^r$, $L_6^r$ and $2 L_1^r - L_2^r$ show no sign of
large-$N_c$ suppression.

In order to proceed, we had to make some assumptions about the NNLO
LECs. For most of the 34 combinations of the $C_i$ appearing in our 17
input variables, theoretical predictions exist in the literature (see 
Ref.~\cite{Bijnens:2014lea}), although in some cases conflicting with
each other. We
have used the available information to define priors for the $C_i$
with associated ranges of acceptable values. We then use an iterative
procedure employing two different methods (minimization or random
walk) in the restricted space of the $C_i$, with essentially the same
results: if the fitted values of the $L_i$ deviate too much from the
NLO fit results (shown in the fourth column of Table \ref{tab:be14})
and/or if the $\chi^2$ is too large, we modify the boundaries of the
$C_i$ space and start again. To speed up the convergence of this
procedure, we introduced a penalty function for bad convergence of the
meson masses. On the other hand, we did not
enforce large $N_c$ on the $C_i$ from the outset.

\begin{center}
\begin{table}[!ht]
\begin{center} 
\begin{tabular}{|c||r|r||r|r|}
\hline
 & NNLO free fit  & \textcolor{red}{NNLO BE14} & NLO 2014 & GL 1985
\cite{Gasser:1984gg} \\  
\hline
& & & & \\[-.3cm] 
$10^3 L_A^r$&     0.68(11) & \textcolor{red}{0.24(11)}  & 0.4(2)  &\\
$10^3 L_1^r$&    0.64(06) & \textcolor{red}{0.53(06)} & 1.0(1)  & 0.7(3) \\
$10^3 L_2^r$&    0.59(04) & \textcolor{red}{0.81(04)} & 1.6(2) & 1.3(7) \\
$10^3 L_3^r$&  $-$2.80(20) & \textcolor{red}{ $-$3.07(20)} &
$-$3.8(3) &  $-$4.4(2.5) \\ 
$10^3 L_4^r$&  0.76(18)  & \textcolor{red}{0.3} & 0.0(3) & $-$0.3(5) \\
$10^3 L_5^r$&  0.50(07) & \textcolor{red}{1.01(06)} & 1.2(1)  & 1.4(5) \\
$10^3 L_6^r$&   0.49(25) &  \textcolor{red}{0.14(05)}& 0.0(4) & $-$0.2(3) \\
$10^3 L_7^r$&  $-$0.19(08) & \textcolor{red}{$-$0.34(09)} & $-$0.3(2) &  $-$0.4(2)\\
$10^3 L_8^r$&   0.17(11) & \textcolor{red}{0.47(10)} & 0.5(2) & 0.9(3) \\
\hline                                                        
$F_0$ [MeV] &    64 & \textcolor{red}{71} &  &
       \\
\hline
\end{tabular}
\end{center} 
\caption{Various fits for the NLO LECs $L_i^r(i=1,\dots,8)$. The values
  for the large-$N_c$ suppressed combination $L_A^r := 2 L_1^r - L_2^r$ 
  are shown separately. We suggest to use fit BE14
  \cite{Bijnens:2014lea} in NNLO
  chiral $SU(3)$ calculations. The last two columns confront an
  up-to-date NLO fit with the original values from
  Ref.~\cite{Gasser:1984gg}.} 
\label{tab:be14}
\end{table}
\end{center}

At the end of the iteration we perform a standard $\chi^2$ fit
for the $L_i$ with fixed ``best'' values of the $C_i$. The results are
shown in Table \ref{tab:be14} in the second and third columns. For
comparison, we also exhibit an NLO fit (4th column) and the original
values of Gasser and Leutwyler \cite{Gasser:1984gg} (last column).

Referring to Ref.~\cite{Bijnens:2014lea} for more details, I include a
few comments here.

\begin{itemize} 
\item All fits are very sensitive to $L_4$ and tend to lead to
  values incompatible with large $N_c$. This tendency is well known
  from previous global fits and can be understood to some extent from
  the structure of the chiral Lagrangian \cite{Ecker:2013pba}. We have
  therefore enforced small values of $L_4$ as supported by lattice
  studies \cite{Aoki:2013ldr}. It turns out that $L_6^r$ and
  $L_A^r := 2 L_1^r - L_2^r$  are then automatically suppressed as well.
\item The quality of both NNLO fits in Table \ref{tab:be14} is
  excellent. They only make sense with associated sets of the $C_i$
  (see Ref.~\cite{Bijnens:2014lea}). 
\item Once a decent convergence is enforced for the meson masses, we
  find reasonable convergence for the other observables. The fitted
  $L_i$ together with the ``best'' values of the $C_i$ show again
  qualitative evidence for resonance saturation
  \cite{Ecker:1988te}. In particular, $\eta^\prime$ dominance is
  observed for some of the $C_i$ in accordance with large $N_c$
  \cite{Kaiser:2007zz}.    
\item Our preferred fit BE14 is remarkably consistent with the NLO
  fits in the last two columns of Table \ref{tab:be14} spanning a
  period of nearly 30 years.
\end{itemize} 

For a determination of other LECs (not covered by our fits) from
$\tau$ decay data I refer to the contribution of Golterman
\cite{marten}. 


\section{Status of CKM unitarity}
\label{sec:ckm}

Violation of unitarity of the three-generation
Cabibbo-Kobayashi-Maskawa mixing matrix would be a clear indication for
physics beyond the Standard Model. The most sensitive test involves
the first row of the CKM matrix with elements $V_{ud}$,
$V_{us}$ and $V_{ub}$. In view of the present sensitivity, only
$V_{ud}$ and $V_{us}$ are relevant for the unitarity test.

Table \ref{tab:ckm} collects the different sources of input for $V_{ud}$ and
$V_{us}$.

\begin{center} 
\begin{table}[!ht]
\begin{center} 
\begin{tabular}{|c|c|c|} \hline
& &
\\[-.3cm] 
matrix element & input source  & significance
 \\[.1cm]  \hline
& &
\\[-.3cm] 
$V_{ud}$  & superallowed $\beta$ decays &
\textcolor{green}{\ding{52}~\ding{52}}  \\[.1cm] 
  & neutron $\beta$ decay & \textcolor{green}{\ding{52}} \\[.1cm]
  & pion $\beta$ decay & 
\\[.1cm]  \hline
$V_{us}$  & $K_{\ell 3}$ decays $\to  |f_+^{K^0\pi^-}(0) V_{us}|$ &
\textcolor{green}{\ding{52}~\ding{52}}  \\[.1cm] 
  &  $\Gamma(K_{\mu 2})$/$\Gamma(\pi_{\mu 2})$, $F_K/F_\pi$, $V_{ud}$
&  \textcolor{green}{\ding{52}~\ding{52}}\\[.1cm] 
  &  $\Gamma(\tau \to X_s \nu)$ vs. $\Gamma(\tau \to X_{\rm 
nonstrange} \nu)$, $V_{ud}$ &   \textcolor{green}{\ding{52}} \\[.1cm] 
  & hyperon decays & 
\\[.1cm]  \hline
\end{tabular}  
\end{center} 
\caption{Sources of input for the determination of the CKM matrix
  elements $V_{ud}$ and $V_{us}$.}
\label{tab:ckm}
\end{table}    
\end{center} 

\begin{center} 
\begin{figure}[!ht]
\begin{center}  
\leavevmode 
\includegraphics[width=11cm]{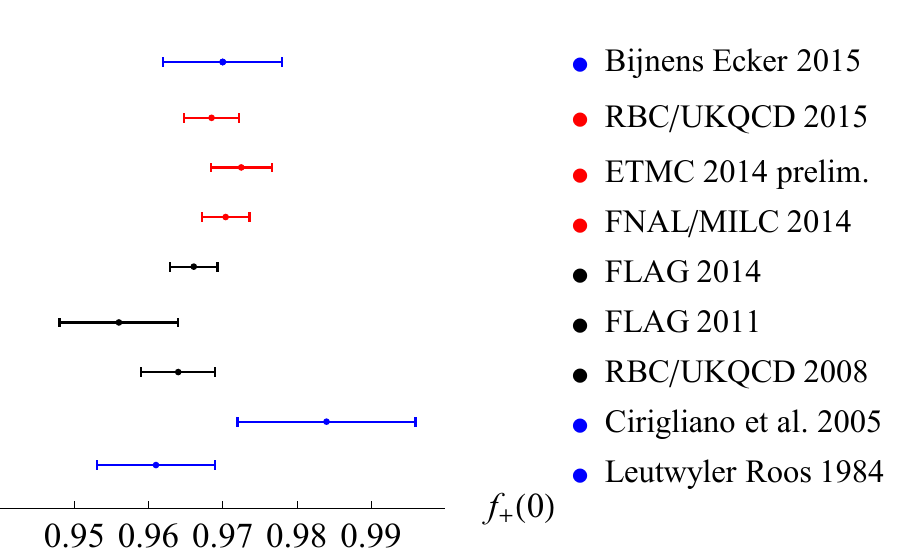}
\end{center} 
\caption{Incomplete list of determinations of $f_+^{K^0\pi^-}(0)$:
  Leutwyler Roos 1984 \cite{Leutwyler:1984je}, Cirigliano et al. 2005
  \cite{Cirigliano:2005xn}, RBC/UKQCD 2008 \cite{Boyle:2007qe}, FLAG 
  2011 \cite{Colangelo:2010et}, FLAG 2014 \cite{Aoki:2013ldr},
  FNAL/MILC 2014 \cite{Bazavov:2013maa}, ETMC (preliminary $SU(3)$
  analysis) \cite{riggio}, 
  RBC/UKQCD 2015 \cite{Boyle:2015hfa}, Bijnens Ecker 2015
  \cite{be2015}. } 
\label{fig:fzero}
\end{figure}
\end{center}

To pin down $V_{us}$ independently of  $V_{ud}$, one needs
$f_+^{K^0\pi^-}(0)$, the  $K_{\ell 3}$ vector form factor at
$t=0$. Some steps in the 30-year-history of estimates/determinations
of $f_+^{K^0\pi^-}(0)$ are listed in Fig.~\ref{fig:fzero}.     

\begin{center} 
\begin{figure}[!ht]
\begin{center}  
\leavevmode 
\includegraphics[width=11cm]{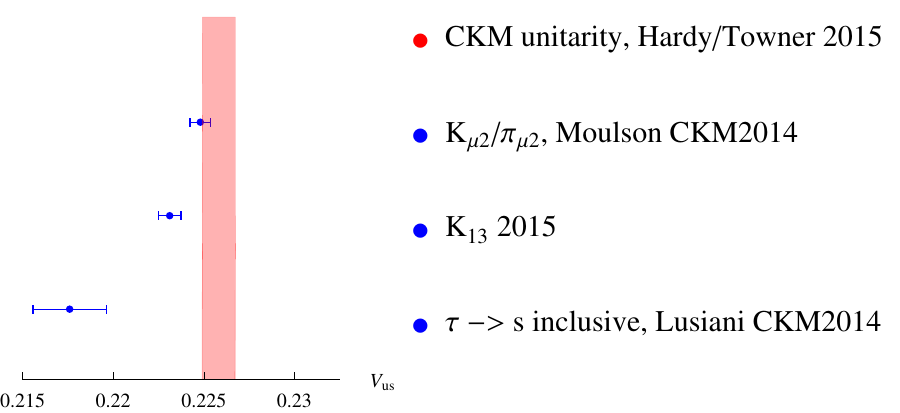}
\end{center} 
\caption{Values for $V_{us}$ using the various inputs discussed in
  the text. The red bar indicates the prediction from unitarity with
  $V_{ud}$ taken from Ref.~\cite{Hardy:2014qxa}.}
\label{fig:plotA}
\end{figure}
\end{center}

The three most recent lattice determinations are marked in red and
will be used for the unitarity test below. Combining the NNLO CHPT
calculation of $K_{\ell 3}$ form factors \cite{Bijnens:2003uy} and fit
BE14 for the LECs \cite{Bijnens:2014lea}, one finds \cite{be2015}
\begin{equation}
f_+^{K^0\pi^-}(0) = 1 - \underbrace{0.02276}_{\rm NLO} -
\underbrace{0.00754}_{\rm NNLO} ~,
\end{equation}
leading to the CHPT prediction
\begin{equation} 
f_+^{K^0\pi^-}(0) = 0.970 \pm 0.008
\end{equation} 
with a very cautious error estimate. This value
should be compared with the average of the three most recent 
lattice determinations (marked in red in Fig.~\ref{fig:fzero})
\begin{equation} 
f_+^{K^0\pi^-}(0) = 0.9703 \pm 0.0021
\label{eq:fzero}
\end{equation} 
to be used for the following unitarity test ($K_{\ell 3}$ 2015
in Fig.~\ref{fig:plotA}). The other input for
Fig.~\ref{fig:plotA} is taken from the contributions of Lusiani
\cite{Lusiani:2014aga} and Moulson \cite{Moulson:2014cra} at CKM2014,
except for the most recent update of $V_{ud}$ \cite{Hardy:2014qxa}.

\begin{center} 
\begin{figure}[!ht] 
\begin{center} 
\leavevmode 
\includegraphics[width=9cm]{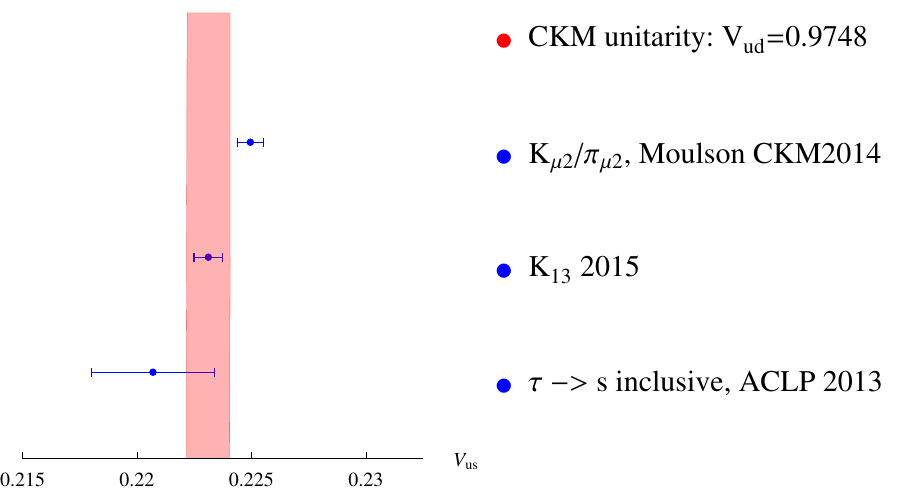}
\end{center} 
\caption{Compared to 
the previous figure, the value for  $V_{us}$
  from $\tau$ decays is now from 
  Antonelli et al. (ACLP 2013) \cite{Antonelli:2013usa} and 
  $V_{ud}=0.97480(21)$ is bigger than the standard value 
  \cite{Hardy:2014qxa} by 3 $\sigma$.} 
\label{fig:plotE}
\end{figure}
\end{center}

Fig.~\ref{fig:plotA} indicates that, instead of worrying about CKM
unitarity, one should first try to straighten out the seemingly
disparate entries for  $V_{us}$. Especially the extraction from $\tau$
data looks problematic compared to the two precise determinations from
$K$ decays. At least among many theorists, the measured
branching ratio $B(\tau \to X_s \nu)$  is often suspected to be the
culprit. Antonelli et al. therefore proposed \cite{Antonelli:2013usa}
to use $K$ decays and lepton universality to calculate the decay widths
$\Gamma(\tau \to K \nu)$ and $\Gamma(\tau \to K \pi \nu)$, which together
make up 68$\%$ of the total strange width. This indeed shifts the
$V_{us}$ entry from $\tau$ decays in the right direction as shown in
Fig.~\ref{fig:plotE}. For the sake of the argument, I have in addition
increased in Fig.~\ref{fig:plotE} the value of $V_{ud}$ from
Ref.~\cite{Hardy:2014qxa} by 3 $\sigma$, anticipating the possible
status of CKM unitarity at the time of Chiral Dynamics 2018 \dots.    


\section{Charged pion polarizabilities}
\label{sec:polar}

\begin{center} 
\begin{figure}[!ht]
\begin{center}  
\leavevmode 
\includegraphics[width=9cm]{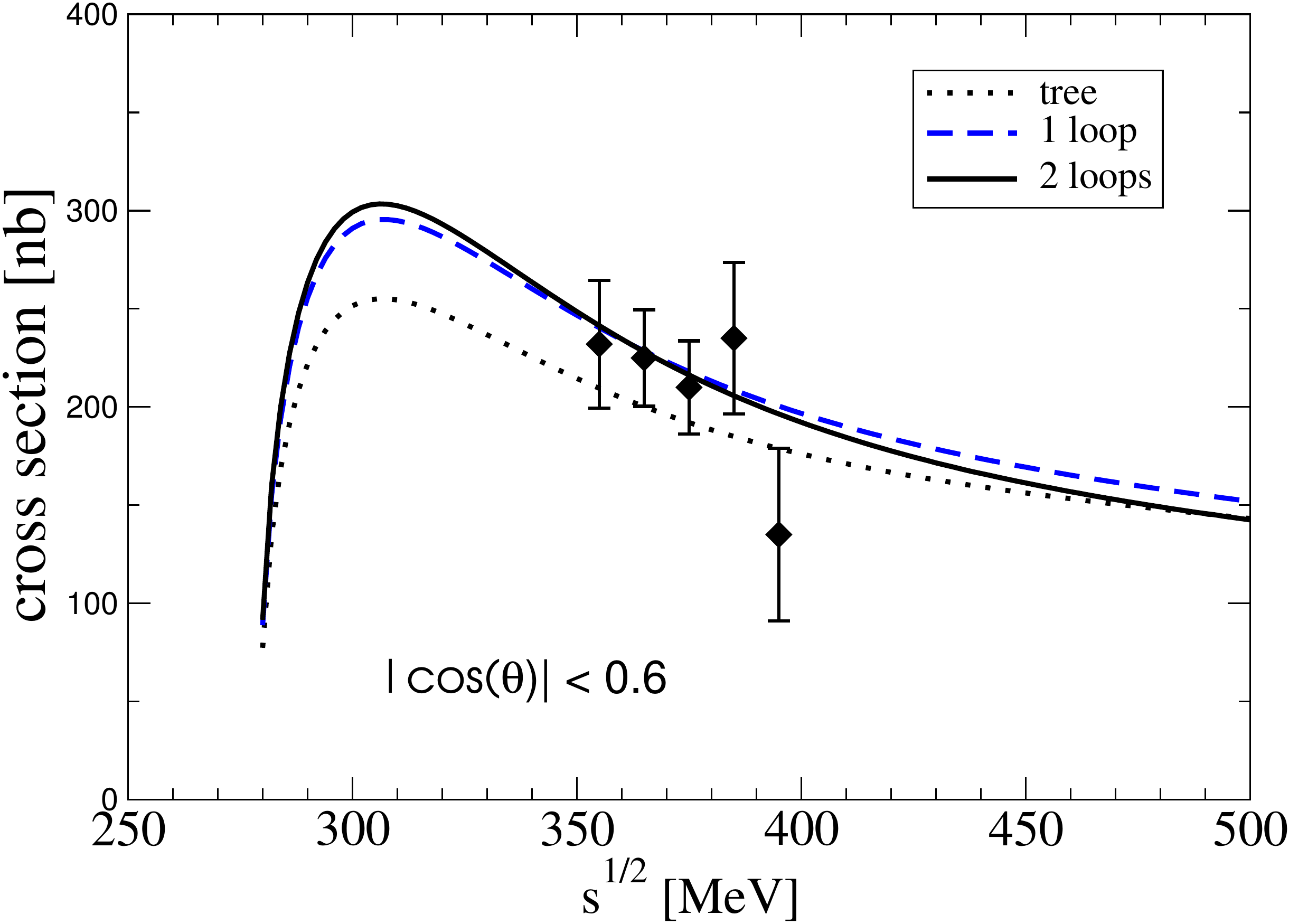}
\end{center} 
\caption{Successive orders of the total cross section $\sigma(\gamma
  \gamma \to \pi^+ \pi^-)$ taken from Ref.~\cite{Gasser:2006qa}.}
\label{fig:ggpipi}
\end{figure}
\end{center}

The $\gamma \gamma \pi^+ \pi^-$ complex is an early example of NNLO
CHPT. Since we are dealing here with chiral $SU(2)$, one expects good
convergence: from scalar electrodynamics at lowest $O(p^2)$ to NLO
\cite{Bijnens:1987dc} and NNLO \cite{Burgi:1996mm,Gasser:2006qa}. This
is indeed borne out by the calculations as exemplified by the total
cross section $\sigma(\gamma \gamma \to \pi^+ \pi^-)$ shown in
Fig.~\ref{fig:ggpipi}.    
The electric ($\alpha_\pi$) and magnetic ($\beta_\pi$)
polarizabilities can be defined via the   
threshold expansion for the differential cross section for pion
Compton scattering $\gamma \pi^\pm \to \gamma \pi^\pm$: 
\begin{equation} 
\displaystyle\frac{d\sigma}{d\Omega} = 
\left(\displaystyle\frac{d\sigma}{d\Omega} \right)_{\rm Born} 
 - \displaystyle\frac{\alpha M_\pi^3(s - M_\pi^2)^2}{4 s^2 (s z_+ +
  M_\pi^2 z_-)} \left(z_-^2 (\alpha_\pi - \beta_\pi ) +
 \frac{s}{M_\pi^4}  z_+^2 (\alpha_\pi + \beta_\pi )  \right) + \dots 
\end{equation} 
with $s=\left(p_{\pi^+} + p_{\pi^-} \right)^2  , ~z_\pm = 1 \pm
\cos{\theta_{\rm cm}}$. 

At NLO in CHPT, electric and magnetic polarizabilities are
equal. In addition to the loop contribution, a single combination of
$SU(2)$ LECs $2 l_5 - l_6$ enters, which is accurately known from
$\pi \to e \nu \gamma$ \cite{Gasser:1983yg}. At NNLO the LECs
$l_1,l_2,l_3,l_4$ (in one-loop diagrams) and three NNLO LECs
contribute together with one-and two-loop contributions. It turns out
that the difference 
$\alpha_\pi - \beta_\pi$ is not very sensitive to the NNLO LECs
leading to the final result\footnote{Polarizabilities are expressed in
units of $10^{-4} {\rm fm}^3$.} $\alpha_\pi - \beta_\pi = 5.7 \pm 1.0$
\cite{Gasser:2006qa}. The sum $\alpha_\pi + \beta_\pi \simeq 0.16$ is
much smaller but the relative uncertainty is bigger than for the
difference. Most experiments actually assume $\alpha_\pi = - \beta_\pi$
in their analyses.

\begin{center} 
\begin{table}[!hb]
\begin{center} 
\begin{tabular}{|c||c|c|} \hline
& &
\\[-.3cm] 
experiment &  & $\alpha_\pi - \beta_\pi$
 \\[.1cm]  \hline
& &
\\[-.3cm] 
MAMI A2 2005 \cite{Ahrens:2004mg}  & $\gamma p \to \gamma \pi^+ n$ &
 $11.6 \pm  1.5_{\rm stat} \pm 3.0_{\rm syst} \pm 0.5_{\rm mod}$  \\[.1cm]
\textcolor{red}{COMPASS 2015  ($\beta_\pi = -
  \alpha_\pi$)} \cite{Adolph:2014kgj}  &  
$\pi^- {\rm Ni} \to \pi^- \gamma ~{\rm Ni}$   &
\textcolor{red}{$4.0 \pm 1.2_{\rm stat} \pm 1.4_{\rm syst}$}
\\[.1cm]  \hline
theory & & 
 \\[.1cm]  \hline
& &
\\[-.3cm] 
Fil'kov, Kashevarov 2006 \cite{Fil'kov:2005ss} & dispersive &
$13.0\left(^{+ 2.6}_{- 1.9}\right)$ \\[.1cm]
\textcolor{red}{Gasser, Ivanov, Sainio 2006} \cite{Gasser:2006qa}
   & CHPT
& \textcolor{red}{$5.7 \pm 1.0$} 
\\[.1cm]  \hline
\end{tabular}  
\end{center} 
\caption{Experimental and theoretical results of the past 10 years for
  the charged pion polarizabilities. }
\label{tab:polar}
\end{table}    
\end{center}

The experimental situation has been inconclusive for a long
time, with large uncertainties ($\alpha_\pi - \beta_\pi \sim ~4
\div  53$). For more details I refer to the plenary talk of Friedrich
\cite{friedrich}. During the last 10 years two more high-precision
experiments were performed. The result of the MAMI experiment of 2005
\cite{Ahrens:2004mg} was incompatible with the chiral prediction for
$\alpha_\pi - \beta_\pi$, but seemed to be supported by a dispersive
analysis\footnote{The dispersive method of Ref.~\cite{Fil'kov:2005ss}
  was criticized by Pasquini et al. \cite{Pasquini:2008ep}.}
soon afterwards \cite{Fil'kov:2005ss}. 

The very recent COMPASS experiment at CERN
\cite{Adolph:2014kgj,friedrich} has given a new twist to this long
story. Their result for  $\alpha_\pi - \beta_\pi$ (assuming
$\alpha_\pi = - \beta_\pi$) displayed in Table \ref{tab:polar} disagrees
with the MAMI result and it is in excellent agreement with the chiral
prediction. The chiral practitioner's favourites
are marked in red in Table \ref{tab:polar}, bringing the longtime
puzzle of the charged pion polarizabilities to a happy ending. It goes
without saying that this success story should be confirmed by an
independent experiment.

\section{Kaon decays}
\label{sec:kaon}

\subsection{Semileptonic $K$ decays}
Since the last Chiral Dynamics Workshop several studies on $K_{\ell 4}$ 
decays have been undertaken within the CHPT framework. In addition to
investigating the $K_{\ell 4}$ form factors and extracting the associated
LECs of chiral $SU(3)$, one gets access to the $\pi\pi$ threshold
region and thus to the $\pi\pi$ scattering lengths. 

For the comparison of theoretical predictions with high-precision
experimental data, isospin-violating corrections became more and more
important. In particular, to determine the $\pi\pi$ scattering lengths
from $K_{e 4}$ data \cite{Batley:2010zza}, it 
is essential to account for mass-difference corrections in 
$\pi\pi$ phase shifts \cite{Colangelo:2008sm}. The $S$-wave scattering
lengths from $K_{e 4}$ data are then in excellent agreement with the
CHPT+Roy equation analysis \cite{Colangelo:2000jc}.   

A complete treatment of isospin violation must include
radiative corrections. Although such corrections are routinely taken
into account in the experimental analysis by means of some Monte Carlo
program, a state-of-the-art CHPT treatment was missing until
recently. Updating and correcting earlier work \cite{Cuplov:2003bj},
Stoffer has now provided the missing link \cite{Stoffer:2013sfa}. In
addition to the complete isospin-violating one-loop corrections for
$K_{\ell 4}$, the calculation is done with an arbitrary cut on the
photon energy for semi-inclusive $K_{\ell 4 \gamma}$ decays. The
complete isospin-violating  
mass effects for form factors are also included, reproducing in
particular the corrections for the phase shifts
\cite{Colangelo:2008sm}. The complete matrix element at NLO including
all isospin-violating corrections of $O(m_u-m_d,\alpha)$ can then be
used by future  $K_{\ell 4}$ experiments in their Monte Carlo
programs. 

In two recent papers \cite{Bernard:2013faa,Bernard:2015vqa,marc},
Bernard, Descotes-Genon and Knecht have extended the analysis of
mass-difference effects for the $\pi\pi$ phase shifts beyond the
one-loop approximation. Their starting point is the observation that
the phase-shift difference measured by NA48/2 \cite{Batley:2010zza}
can be related to the theoretical expression \cite{Colangelo:2000jc}
as 
\begin{equation} 
\left[\delta_S(s) - \delta_P(s)\right]_{\rm exp} = 
\delta_{\rm Roy}^{S-P}(s;a_0^0,a_0^2)  + \delta_{\rm IB}^{L=1}(s)
\label{eq:delta}
\end{equation} 
where $\delta_{\rm IB}^{L=1}(s)$ is the one-loop correction 
\cite{Colangelo:2008sm}. Beyond one loop, $\delta_{\rm IB}(s)$
will depend on the scattering lengths $a_0^0,a_0^2$ in a nontrivial
way so that 
Eq.~(\ref{eq:delta}) should in general be replaced by
\begin{equation} 
\left[\delta_S(s) - \delta_P(s)\right]_{\rm exp} = 
\delta_{\rm Roy}^{S-P}(s;a_0^0,a_0^2)  +
\delta_{\rm IB}(s;a_0^0,a_0^2)~.
\end{equation} 

The dependence of $\delta_{\rm IB}(s;a_0^0,a_0^2)$ on the scattering
lengths could therefore introduce a bias in the extraction of
$a_0^0,a_0^2$ from the data. Of course, in the usual chiral expansion
this dependence will be hidden in loop contributions and
LECs. Therefore, Bernard et al. set up a dispersive representation of 
$\delta_{\rm IB}(s;a_0^0,a_0^2)$ consisting of two parts
\cite{Bernard:2013faa}: a universal 
part involving only $\pi\pi$ rescattering and a process-dependent part
involving the form factors in the coupled channels. Refitting the
NA48/2 data for $K^\pm \to \pi^+ \pi^- e^\pm \nu_e$
\cite{Batley:2010zza} with this dispersive representation,
the two contributions partially cancel to give rise to 
isospin-breaking corrections close to the one-loop case as shown in
Table \ref{tab:a0a2}.

\begin{center} 
\begin{table}[!ht]
\begin{center} 
\begin{tabular}{|c|c|c|} \hline
& &
\\[-.3cm] 
$a_0^0$ & $a_0^2$  & Ref.
 \\[.1cm]  \hline
& &
\\[-.3cm] 
$0.221 \pm 0.018$  & $- 0.0453 \pm 0.0106$ & \cite{Bernard:2013faa}
\\[.2cm]  
\hspace*{0.4cm} $0.2220(128)_{\rm stat}(50)_{\rm syst}(37)_{\rm th}$ 
\hspace*{0.4cm}  &  
\hspace*{0.4cm} $ - 0.0432(86)_{\rm stat}(34)_{\rm syst}(28)_{\rm th}$ 
\hspace*{0.4cm} & \cite{Batley:2010zza} 
\\[.1cm]  \hline
\end{tabular}  
\end{center} 
\caption{$S$-wave scattering lengths extracted from the NA48/2 data
  for  $K^\pm \to \pi^+ \pi^- e^\pm \nu_e$ \cite{Batley:2010zza}. The
entries in the second line are based on the dispersive analysis of
Ref.~\cite{Bernard:2013faa}, those in the third line are from the
NA48/2 Collaboration \cite{Batley:2010zza} with one-loop
isospin-breaking corrections \cite{Colangelo:2008sm} applied.}
\label{tab:a0a2}
\end{table}    
\end{center} 

\begin{center} 
\begin{figure}[!ht]
\begin{center}  
\leavevmode 
\includegraphics[width=10cm]{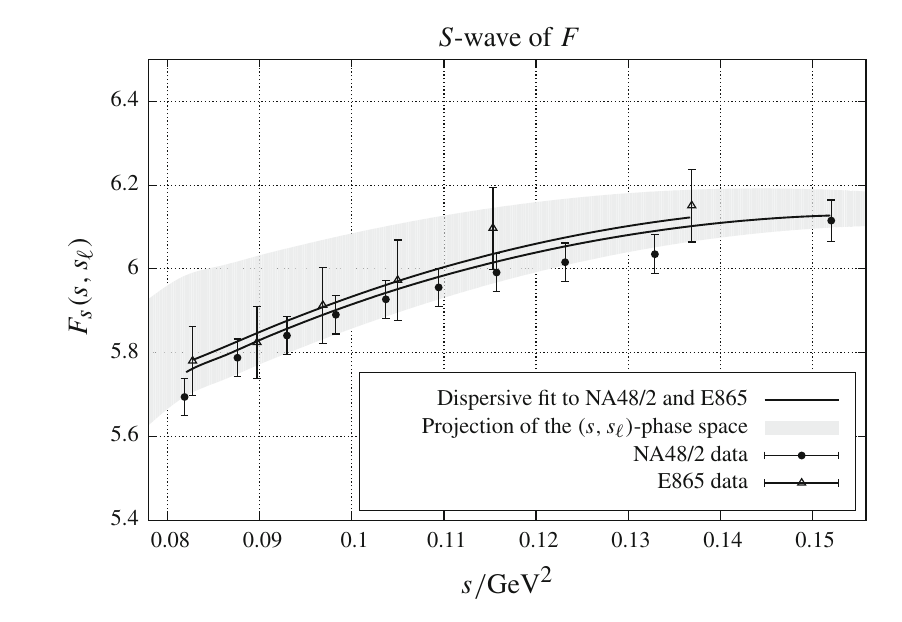}
\end{center} 
\caption{$S$-wave of the $K_{\ell 4}$ form factor $F$. The dispersive
  description \cite{Colangelo:2015kha} accounts for the measured
  curvature of the form factor.} 
\label{fig:Fs}
\end{figure}
\end{center}

Very recently, Colangelo, Passemar and Stoffer have extended the NNLO
calculation of $K_{\ell 4}$ decay amplitudes \cite{Amoros:2000mc} by a
dispersive treatment including resummation of $\pi\pi$ and $K\pi$
rescattering effects \cite{Colangelo:2015kha}, thereby improving the
perturbative treatment of $\pi\pi$ final-state interactions. With
the usual assumption of two-particle rescattering with $S$- and
$P$-waves only (``reconstruction theorem'' \cite{Stern:1993rg}), they
determine (most of) the subtraction constants by fitting to the data of
the high-statistics experiments E865 \cite{Pislak:2003sv} and 
NA48/2 \cite{Batley:2012rf}. Isospin breaking is taken into account 
\cite{Stoffer:2013sfa}. By matching to CHPT at both the one- and the
two-loop 
level, values for the NLO LECs $L_1,L_2,L_3$ are determined that are
compatible with the global fit BE14 \cite{Bijnens:2014lea}. Unlike the
CHPT calculation to NNLO, the dispersive treatment can account for the
measured curvature of the $S$-wave projection of the form factor $F$
as shown in Fig.~\ref{fig:Fs}.

\subsection{Nonleptonic $K$ decays}
Since the last Chiral Dynamics Workshop in 2012, there have been no
spectacular developments in the CHPT treatment of nonleptonic kaon
decays to the best of my knowledge. In this respect, the review of
kaon physics in the 
Standard Model by Cirigliano et al. \cite{Cirigliano:2011ny} is
therefore still up-to-date.

Let me nevertheless mention two recent investigations of nonleptonic
$K$ decays discussed at this Workshop even though they are not directly
related to CHPT. Garron \cite{garron} presented the first complete
lattice calculation of $K \to \pi\pi$ with physical kinematics,
opening the way for an ab-initio verification of the $\Delta I=1/2$
rule and for extracting the parameter $\epsilon^\prime$ measuring direct CP
violation \cite{Bai:2015nea}. The lattice value for
$\epsilon^\prime/\epsilon$ comes out substantially smaller than the
experimental value \cite{Agashe:2014kda} but isospin-violating
corrections \cite{Cirigliano:2003nn,Cirigliano:2003gt}
still need to be applied.

A completely different, if somewhat
unconventional explanation of the $\Delta I=1/2$ rule has been
proposed by Crewther and Tunstall \cite{Crewther:2013vea,tunstall}
assuming a QCD infrared fixed point.

\section{$\eta \to 3 \pi$ decays}
\label{sec:eta}

The decays $\eta \to \pi^+\pi^-\pi^0, 3 \pi^0$ violate isospin. 
Electromagnetic contributions are known to be small
\cite{Sutherland:1966zz,Bell:1996mi}, but they can be incorporated
consistently \cite{Baur:1995gc,Ditsche:2008cq}. To a very good
approximation, the decay amplitudes are therefore proportional to the
quark mass difference $m_d - m_u$:
\begin{eqnarray} 
& ~A(\eta \to 3 \pi)  ~\sim m_d -m_u ~\sim R^{-1} ~\sim Q^{-2} 
\label{eq:defRQ} ~, \\[.1cm] 
& ~R = \displaystyle\frac{m_s - m_{ud}}{m_d - m_u}, ~~Q^2 =
\displaystyle\frac{m_s^2 - m_{ud}^2}{m_d^2 - m_u^2} ~, \no \\[.1cm] 
& ~Q^2 = \left(1 +\frac{m_s}{m_{ud}}\right) R/2,~~m_{ud} := (m_u +
m_d)/2 ~. \no
\end{eqnarray}    

\begin{center} 
\begin{figure}[!ht]
\begin{center}  
\leavevmode 
\includegraphics[width=11cm]{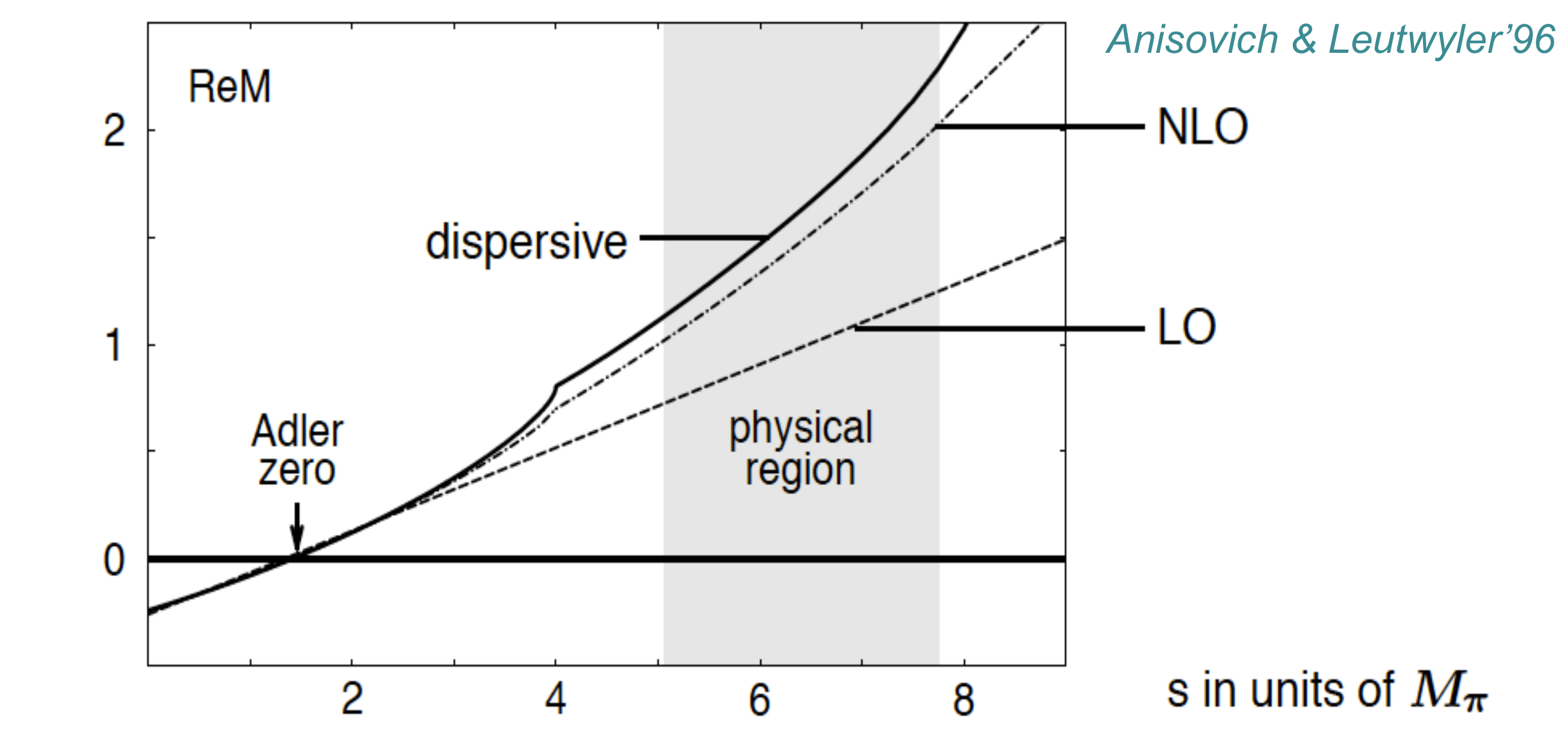}
\end{center} 
\caption{Real part of the (properly normalized)
  amplitude $M(\eta \to \pi^+ \pi^- \pi^0)$ 
  along the line $s=u$: LO (dashed curve), NLO (dash-dotted) and 
  dispersive amplitude (full curve). The figure was taken from 
  Ref.~\cite{Anisovich:1996tx}.} 
\label{fig:anileut}
\end{figure}
\end{center}

From the chiral point-of-view, $\eta \to 3 \pi$ decays are therefore a
unique source of information for extracting the light quark mass
difference $m_d - m_u$, little affected by electromagnetic
effects. Unfortunately, successive orders in the chiral expansion 
do not show any sign of convergence as displayed in Table
\ref{tab:etarate}.  
\begin{center}
\begin{table}[!h]
\begin{center}  
\begin{tabular}{|c||c|c|c|} \hline
& & &
\\[-.3cm] 
&  & $\Gamma(\eta \to \pi^+ \pi^- \pi^0)$/eV & $r$   \\[.1cm]
\hline 
LO & Osborn, Wallace 1970 \cite{Osborn:1970nn} & $66^*$  & $1.54$
 \\[.1cm]  
NLO  & Gasser, Leutwyler 1985 \cite{Gasser:1984pr} & $160(50)^*$ &
$1.46$  \\[.1cm]  
NNLO &  Bijnens, Ghorbani 2007 \cite{Bijnens:2007pr} &  & 1.47 \\[.1cm]
expt. &  PDG 2014 \cite{Agashe:2014kda} &  300(12) & $1.48(5)$
\\[.1cm]  \hline
\end{tabular}
\end{center} 
\caption{Successive orders in the chiral expansion for the decay rate
  $\Gamma(\eta \to \pi^+ \pi^- \pi^0)$ and for the ratio
  $r=\Gamma(\eta \to 3 \pi^0)/\Gamma(\eta \to \pi^+ \pi^- \pi^0)$ in
  comparison with experiment. For the numbers with an asterisk
  $Q=24.3$ (Dashen's theorem \cite{Dashen:1969eg}) has been 
  assumed.} 
\label{tab:etarate}
\end{table}  
\end{center} 
In addition, the CHPT amplitudes have problems with the measured
Dalitz plot distributions as will be discussed below. It has long been
suggested that these problems have to do with the fact that
$\pi\pi$ rescattering is included only perturbatively, suggesting a
dispersive treatment \cite{Kambor:1995yc,Anisovich:1996tx}. Fig.~
\ref{fig:anileut} taken from Ref.~\cite{Anisovich:1996tx} nicely
illustrates the situation for the real part of the decay amplitude:
while the chiral corrections from LO to NLO are large, the dispersive
effects are actually rather moderate. 
Fig.~\ref{fig:anileut} also indicates 
that the decay amplitudes (with or without dispersive
corrections) deviate only very little from the Adler zero \cite{Adler:1964um} 
at $s=u=4/3 M_\pi^2$ as it should be for a chiral $SU(2)$ prediction.

More recent developments to improve the chiral amplitudes with
dispersion theoretic methods are all (with the exception of a recent
attempt to include also resonance effects \cite{bachir}) based on the
so-called 
reconstruction theorem \cite{Stern:1993rg}, incorporating $\pi\pi$
partial-wave discontinuities for $\ell=0,1$ only. The dispersive
amplitudes are then matched to the CHPT amplitudes (mostly at NLO) to
obtain the ratios $Q$ or $R$ from the experimental rates. 
Since the rates cannot be predicted in the dispersive approaches, the
crucial tests involve the Dalitz plot parameters defined in
Eqs.~(\ref{eq:dalitzpar},\ref{eq:dalitzdefs}) through an expansion
around the center of the Dalitz plot for both charged and neutral
modes, especially in view of recent very precise experimental results.
In most of the cases to be discussed below, some of the Dalitz plot
parameters are also used as input.

The Dalitz plot parameters $a,b,d,\alpha$ are defined via the
expansions

\begin{eqnarray} 
|A_{\rm charged}(s,t,u)|^2  &=& N \left(1+ a Y + b Y^2 + d X^2 +
\dots  \right) ~, \no \\[.1cm] 
|A_{\rm neutral}(s,t,u)|^2  &=& N \left(1+ 2 \alpha Z  +\dots \right)
\label{eq:dalitzpar}
\end{eqnarray}  
where
\begin{eqnarray} 
& X=\displaystyle\frac{\sqrt{3}}{2 M_\eta Q_\eta} (u-t), ~
Y=\displaystyle\frac{3}{2 M_\eta Q_\eta} \left((M_\eta - M_{\pi^0})^2 -s
\right) -1~, \label{eq:dalitzdefs} \\[.1cm]
& Z = (X^2 + Y^2),~ Q_\eta = M_\eta - \sum_i M_{\pi^i}.  \no
\end{eqnarray}

During the past years, four groups have employed dispersive
approaches to analyze $\eta \to 3 \pi$ decays.

\begin{enumerate} 
\item Colangelo, Lanz, Leutwyler and Passemar
  \cite{Colangelo:2011zz,Leutwyler:2013wna} \\[.2cm] 
This is a modern update of the approach of Anisovich and Leutwyler
\cite{Anisovich:1996tx}, with final results still pending. Some of the
crucial features are:

\begin{itemize} 
\item The Bern $\pi\pi$ phase shifts
  \cite{Ananthanarayan:2000ht,Colangelo:2001df} are used with 
  effectively six subtraction constants, which are constrained by data
  in the charged decay channel.
\item Electromagnetic effects are fully taken into account to NLO
  \cite{Ditsche:2008cq}.  
\item To pin down the absolute magnitude, the dispersive amplitude is
  matched to the chiral NLO amplitude for small values of $s,t,u$.
  The amplitude is compatible with the Adler zero.  
\end{itemize} 
\item Schneider, Kubis and Ditsche \cite{Schneider:2010hs} \\[.2cm] 
 These authors use a nonrelativistic effective field theory to
 two-loop accuracy that
\begin{itemize} 
\item is well suited to study the dynamics of the final-state
  interaction and includes all isospin-violating corrections.
\item It yields relations between charged and neutral Dalitz plots.
\item The rescattering effects lead to sizable corrections for the
  Dalitz plot parameters. The approach
  offers an explanation why NNLO CHPT misses an important
  contribution to the neutral Dalitz parameter $\alpha$ defined in
  Eq.~(\ref{eq:dalitzpar}). 
\end{itemize}
\item Kampf, Knecht, Novotny and Zdrahal \cite{Kampf:2011wr} \\[.2cm]   
This approach uses an analytic two-loop representation with the
following main features.
\begin{itemize} 
\item The amplitudes contain six parameters 
  that are fitted to the Dalitz plot distribution of the charged decay
  mode. 
\item With these constants determined, the authors predict the 
  parameter $\alpha$ for the neutral mode. 
\item They match their dispersive amplitude to the absorptive part of the
  NNLO chiral amplitude in a region where the differences between NLO
  and NNLO amplitudes are small. In fact, this is so far the only
  attempt to match to CHPT at the NNLO level.
\end{itemize} 
The result has been criticized by the Bern group
\cite{Colangelo:2011zz,Leutwyler:2013wna} because the fitted 
amplitude is quite far away from the Adler zero.
\item Guo et al. (JPAC) \cite{Guo:2015zqa} \\[.2cm] 
This is the most recent dispersive analysis.
\begin{itemize} 
\item The dispersive amplitude uses the Madrid/Cracow $\pi\pi$
  phase shifts \cite{GarciaMartin:2011cn} with only three subtraction
  constants. 
\item They fit to the experimental Dalitz plot for the charged mode
  from the WASA/COSY Collaboration \cite{Adlarson:2014aks} and
  then predict the neutral Dalitz parameter $\alpha$.
\item The absolute scale is fixed by matching to NLO CHPT near the
  Adler zero to extract a value for $Q$.
\end{itemize} 
\end{enumerate}

Three high-precision experiments have recently measured the Dalitz
plot parameters, with the greatest accuracy provided by the
new KLOE results presented at this Workshop \cite{simona}. In Table
\ref{tab:dalitz} I confront the experimental results with theoretical
predictions, as far as available. The experimental results
for the Dalitz plot parameters $a,b,d$ in the charged channel are
consistent with each other and can therefore be averaged. The averages
are clearly dominated by the 
KLOE results. Comparing with theoretical predictions, the least one
can say is that the data present challenges for most of the theoretical
approaches. In particular, NNLO CHPT has serious problems with the
parameters $b$ and $\alpha$. It remains to be seen whether the new set
of LECs BE14 can improve the situation.  
     
\begin{center} 
\begin{table}[!ht]
\begin{center}  
\begin{tabular}{|c||c|c|c||c|} \hline
& & & &
\\[-.3cm] 
& $-a$ & $b$ & $d$ & $\alpha$   \\[.1cm]
\hline 
WASA/COSY 2014 \cite{Adlarson:2014aks} & $1.144(18)$ & $0.219(19)(47)$ &
$0.086(18)(15)$ & \\[.1cm] 
BESIII 2015 \cite{Ablikim:2015cmz} & $1.128(15)(8)$ &
$0.153(17)(4)$ & $0.085(16)(9)$  &   $- 0.055(14)(4)$  
 \\[.1cm]  
KLOE 2015 \cite{simona} & $1.095(3)(^{+3} _{-2})$ &
$0.145(3)(5)$ & $0.081(3)(^{+6} _{-5})$  &  
 \\[.1cm] 
my averages & $1.099(4)$ & $0.147(6)$ & $0.082(6)$ & \\[.1cm]
PDG 2014 \cite{Agashe:2014kda} & & & & $- 0.0315(15)$
\\[.1cm]   
\hline 
NNLO CHPT \cite{Bijnens:2007pr} & $1.271(75)$ & $0.394(102)$ & $0.055(57)$
& $0.013(32)$  \\[.1cm] 
NREFT \cite{Schneider:2010hs} & $1.213(14)$ & $0.308(23)$ & $0.050(3)$
& $-0.025(5)$  \\[.1cm]   
KKNZ \cite{Kampf:2011wr} & & & & $- 0.044(4)$ \\[.1cm] 
JPAC \cite{Guo:2015zqa} & $1.116(32)$ & $0.188(12)$ & $0.063(4)$ & $-
0.022(4)$  \\[.1cm]
\hline  
\end{tabular}
\end{center} 
\caption{Experimental data from three recent high-precision
  experiments on the Dalitz plot parameters (\protect\ref{eq:dalitzpar})
  (charged and neutral modes). In the lower half of the table, the
  available theoretical results are shown for comparison. }
\label{tab:dalitz}
\end{table}    
\end{center} 

Finally, one can extract values for the quark mass ratios from the
experimental rates. In Table 
\ref{tab:RQ} I collect the values for the ratios $R$ and $Q$
(\ref{eq:defRQ}) together with $N_f=2+1$ lattice averages
\cite{Aoki:2013ldr}. The general tendency can be summarized in the
following way. Since the Prague/Marseille group  \cite{Kampf:2011wr}
matches to NNLO CHPT, their values for  $R,Q$ are bigger than
those from Refs.~\cite{Colangelo:2011zz,Guo:2015zqa} where the
matching was performed at the NLO level. The lattice values
\cite{Aoki:2013ldr} are in between.

\begin{center} 
\begin{table}
\begin{center}  
\begin{tabular}{|c|c|c|c|c||c|} \hline
& & & & &
\\[-.3cm] 
& NNLO CHPT \cite{Bijnens:2007pr,JBprivate} & CLLP \cite{Colangelo:2011zz} & 
KKNZ \cite{Kampf:2011wr,marcprivate} & JPAC \cite{Guo:2015zqa} & FLAG 2014
\cite{Aoki:2013ldr}  \\[.1cm]  
& &  prel. & & & ($N_f=2+1$)
\\[.1cm]
\hline 
& & & & &
\\[-.3cm] 
$R$ &  $40.9$ & $[31.9]^*$ & $37.4(2.2)$ & $[32.2]^*$  &
$35.8(1.9)(1.8)$\\[.1cm] 
$Q$ & $[24.1]^*$ & $21.3(6)$ & $[23.1]^*$ & $21.4(4)$ & $22.6(7)(6)$
\\[.1cm]  \hline
\end{tabular}
\end{center} 
\caption{Theoretical predictions for the quark charge ratios $R,Q$.
 The values without
  brackets are taken directly from the publications.  The values in
  brackets with an asterisk (without errors) were calculated using
  the relation between $R$ and $Q$ in Eq.~(\protect\ref{eq:defRQ})
  and the FLAG value \cite{Aoki:2013ldr} for  $m_s/m_{ud}$ .} 
\label{tab:RQ}
\end{table}    
\end{center}


\section{Conclusions}

Chiral perturbation theory for light mesons is still strong and
healthy after more than 30 years. The main challenge for CHPT to
understand the physics of the Standard Model in the confinement regime
has been met successfully. This applies especially to chiral $SU(2)$
where the longtime puzzle of the charged pion polarizabilities has
now been resolved, with an impressive agreement between CHPT and
experiment. 

With chiral $SU(3)$, there is still ample room for improvements. The
abundance of low-energy constants at NNLO is one major obstacle on the
way to precision physics. A new global fit for the LECs in the strong
sector is a promising step towards improving the situation.
With the new set of LECs, NNLO CHPT and the most recent lattice
results for the $K_{\ell 3}$ vector form factor at $t=0$ are now in
excellent agreement, leading to a precise value of $V_{us}$. 

The second main construction site for chiral $SU(3)$ is the
slow convergence of the chiral series in some cases. The problem
has been attacked by several groups with dispersive methods, both for  
$K_{\ell 4}$ decays and especially for $\eta \to 3 \pi$ where
CHPT seems unable to account for the measured Dalitz plot
distributions even at NNLO. Although the dispersive approaches depend
heavily on experimental input for pinning down the subtraction
constants, they can incorporate final-state interactions more
effectively than CHPT. In any case, the new very precise data for the
$\eta \to 3 \pi$ Dalitz plot parameters constitute a challenge for all
approaches.  

The lack of evidence for new physics in the low-energy regime should
not be held against CHPT. After all, CHPT is in good company with LHC
physics in this respect.

\acknowledgments
I thank Hans Bijnens for allowing me to present our unpublished
result for the $K_{\ell 3}$ form factor at $t=0$ and Simona Giovanella
for sending me the new KLOE data for the Dalitz plot parameters prior
to the meeting. Last but not least, my thanks are due to Laura
Marcucci and Michele Viviani for a very inspiring and enjoyable Workshop.

\end{document}